\documentstyle[times,graphics,astrobib,amssymb,epsfig,subfigure]{mn2e}

\def\labfig #1{\label{fig:#1}}
\def\labsecn #1{\label{sec:#1}}
\def\labtablem #1{\label{tab:#1}}

\def\fig #1{Figure~\ref{fig:#1}}
\def\secn #1{Section~\ref{sec:#1}}
\def\tablem #1{Table~\ref{tab:#1}}

\def\unit #1{\,{\rm #1}}
\def\nh{N_{\rm H}}
\def\cmsqi{\unit{cm^{-2}}}
\def\kel{\unit{K}}
\def\etal{et al.\ }
\def\onlyten#1{10^{#1}}
\def\cmcui{\unit{cm^{-3}}}
\def\ev{\unit{eV}}
\def\kev{\unit{keV}}

\title[DR \& warm gas in AGN]
{Dielectronic recombination and stability of warm gas in AGN}
\author[Chakravorty \etal]
{Susmita Chakravorty$^{1}$
\thanks{E-mail: susmita@iucaa.ernet.in (SC); akk@iucaa.ernet.in (AK); elvis@head.cfa.harvard.edu (MA); gary@pa.uky.edu (GF); badnell@phys.strath.ac.uk (NB)},
Ajit K. Kembhavi$^{1*}$, Martin Elvis$^{2*}$, Gary Ferland$^{3*}$, N.R.Badnell$^{4*}$ \\
$^{1}$IUCAA, Post Bag 4, Ganeshkhind, Pune 411 007, India; \\
$^{2}$Harvard-Smithsonian Center for Astrophysics, Cambridge, MA 02138; \\
$^{3}$Department of Physics and Astronomy, University of Kentucky, Lexington, KY 40506; \\
$^{4}$Department of Physics, University of Strathclyde, Glasgow G4 0NG, UK. }

\begin{document}

\maketitle


\begin{abstract}

Understanding the thermal equilibrium (stability) curve may offer insights
into the nature of the warm absorbers often found in active galactic nuclei.
Its shape is determined by factors like
the spectrum of the ionizing continuum and the chemical
composition of the gas. We find that the stability curves obtained
under the same set of the above mentioned physical factors, but using recently derived
dielectronic recombination rates, give significantly
different results, especially in the regions corresponding to
warm absorbers, leading to different physical predictions.
Using the current rates we find a larger probability of having
thermally stable warm absorber at $10^5 \kel$ than previous
predictions and also a greater possibility for its multiphase
nature. the results obtained with the current dielectronic
recombination rate coefficients are more reliable because
the warm absorber models along the stability curve have
computed coefficient values, whereas previous calculations relied on
guessed averages for the same due to lack of available data.

\end{abstract}




\section{Introduction}
\labsecn{introduction}

Warm Absorbers are highly photoionised gas found along our line
of sight to the continuum source of active galactic nuclei (AGN).
Their signatures are a wealth of absorption lines and edges from
highly ionized species, notably OVII, OVIII, FeXVII, NeX, CV
and CVI in the soft X-ray (0.3-1.5 keV) spectra. The typical
column density observed for the gas is $\nh \sim
10^{22\pm1}\cmsqi$ and the temperature $T$ is estimated to be a
few times $10^5\kel$. For many objects the warm absorber exists
as a multiphase absorbing medium with all the phases in near
pressure equilibrium (Morales, Fabian \& Reynolds, 2000; Collinge
\etal 2001; Kaastra \etal 2002; Krongold \etal 2003; Netzer \etal
2003; Krongold \etal 2005; Ashton \etal 2006).

Any stable photoionised gas will lie on the thermal equilibrium curve
or `stability' curve, in the temperature - pressure phase space,
where heating balances cooling;
this curve is often used to study the multiphase nature of the
photoionised gas. The equilibrium depends on the shape of the ionizing
continuum and the chemical abundance of the gas
(Krolik, McKee \& Tarter, 1981; Krolik \& Kriss, 2001;
Reynolds \& Fabian, 1995 and Komossa \& Mathur, 2001). We are investigating
this dependences in details and will report elsewhere
(Chakravorty \etal, to be submitted).

Over the past two decades the estimates of dielectronic recombination rate
coefficients have improved.
In this letter we show that these have affected the stability
curves significantly which may lead to quite different physical models
for the warm absorber gas.
We conclude with a caution on the reliability of
earlier results.


\section{The Stability Curve}
\labsecn{scurve}

As is customary, we consider an optically thin, plane parallel slab of
gas being photoionised by the central source of the AGN. In ionization
equilibrium, photoionisation is balanced by recombinations. Thermal
equilibrium is achieved when photoionisation and
Compton heating are balanced by collisional cooling, recombination,
line-excitation, bremsstrahlung and Compton cooling.
The conditions under which these  equilibriums are
achieved depend on the shape of the continuum, metallicity of the gas,
the density, the column density, and the ratio of
the ionizing photon flux to the gas density. Following the convention of
Tarter,
Tucker \& Salpeter (1969) we specify this ratio through an
{\it ionization parameter} $\xi=L/nR^2$, where $L$ is the
luminosity of the source and $n$ the number density of gas at a distance $R$
from the center of the AGN.

We consider a sequence of models for a range of
ionization parameters $\xi$, for optically thin gas having
constant density of $\onlyten{9}\cmcui$, being irradiated by a power-law
ionizing continuum with photon index $\Gamma=1.8$,
so that $f(\nu) \sim \nu^{-(\Gamma-1)}$, and
extending from 13.6 eV to 40 keV. The chemical composition of
the gas (referred to as {\it old Solar abundance}) is
is due to Grevesse and Anders (1989)
with extensions by Grevesse and Noels (1993). These parameter values
(hereafter called the {\it standard set} of parameters) have been
considered because that allows us to compare our results easily
with earlier work.

Using version C07.02.01\footnote{http://www.nublado.org/} (hereafter C07)
of the photoionisation code CLOUDY (see Ferland \etal 1998 for a
description), we plot the thermal stability curve which is shown as
the solid curve (labeled C07) in \fig{stability}. $\xi/T$ is
proportional to $p_{rad}/p$, where $p_{rad}$ is the
radiation pressure and $p$ the gas pressure; so an isobaric
perturbation of a system in equilibrium is represented by a
vertical displacement from the curve, and only
changes the temperature.
If the system being perturbed is located  on a part
of the curve with positive slope, which covers most of the curve,
then a perturbation
corresponding to an increase in temperature leads to cooling,
while a decrease in temperature leads to heating of the gas.
Such a gas is therefore thermally stable. But if the system is
located on one of the few parts of the curve with negative slope,
then it is
thermally unstable because
isobaric perturbations will lead to runaway heating or cooling.


\begin{figure}
\centerline{\psfig{figure=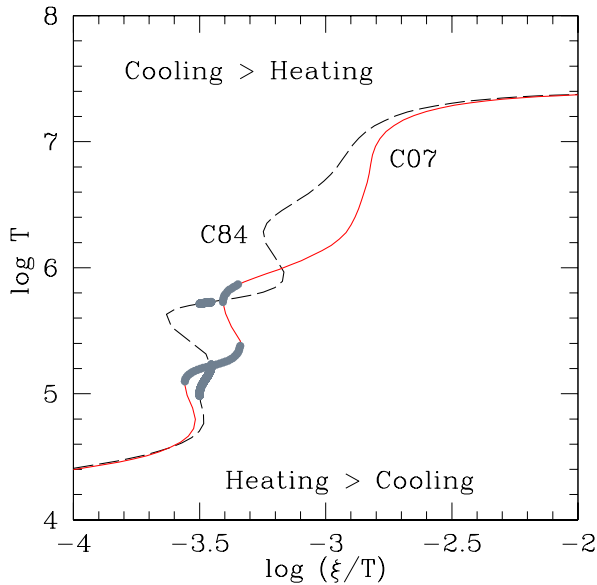,width=8.0cm,angle=0.}}
\caption{Stability curves generated by versions C07 and C84 of CLOUDY,
for an optically thin gas with constant density $\onlyten{9}\cmcui$, being ionized
by a power-law continuum with photon index $\Gamma=1.8$, extending
from $13.6\ev$ to $40\kev$. The gas, in both cases, is assumed to have
{\it old Solar abundance} (see text).
Regions of the plane where heating or cooling
dominates are indicated. The curves are seen to differ significantly in the
temperature range of $4.2 \le \log T \le 7.0$. Multiphase regions on the solid and dashed
curve have been highlighted.}
\labfig{stability}
\end{figure}


\section{Comparison With Previous Work}
\labsecn{compare}

Reynolds \& Fabian (1995, hereafter RF95), have studied the
stability curve for the warm absorber in MCG-6-30-15, using the
standard set of parameters given above.  We have reproduced their
curve, as given in Figure 3 of RF95, using version C84.12a
(hereafter C84) of CLOUDY, which was the stable version between
1993 and 1996; this is shown as the dashed curve in
\fig{stability}. The C84 and C07 curves match at high
temperatures in the range $T>10^{7.2}\kel$, where Compton heating
and cooling dominate. They also agree at low temperatures 
$T\lesssim10^{4.5}\kel$. However, the curves are significantly
different in the intermediate temperature range $\onlyten{4.5}\le
T\le\onlyten{7.2}\kel$, which is the region of interest for the
warm absorbers and where recombination and line excitation are
dominant cooling mechanisms.


\begin{table}
\begin{center}
\begin{tabular}{c c c c c c}
\hline
& & & \multicolumn{2}{c}{$\Delta\log(\xi/T)$} & \\ \cline{4-5}
\raisebox{1.5ex}[0cm][0cm]{Version} & \raisebox{1.5ex}[0cm][0cm]{$\xi_5$} %
& \raisebox{1.5ex}[0cm][0cm]{$N_{\rm phases}$} & %
\raisebox{-0.5ex}[0cm][0cm]{$\sim10^5\kel$} & %
\raisebox{-0.5ex}[0cm][0cm]{$\sim10^6\kel$} & %
\raisebox{1.5ex}[0cm][0cm]{$\Delta_{\rm{M}}\log(\xi/T)$} \\
\hline
C84  & 45 & 2 & 0.05 & 0.47 & 0.05 \\ \\
C07  & 74 & 2 & 0.22 & 0.46 & 0.07 \\
\hline
\end{tabular}
\caption{Parameters for the warm absorber obtained using versions
C84 and C07 of CLOUDY. The second column gives the value of the
ionization parameter of the $\sim10^5\kel$ stable warm absorber phase,
the third column the number of discrete phases in the multiphase medium,
the fourth and fifth
columns the range of $\log(\xi/T)$ respectively for the $\sim10^5\kel$ and
$\sim10^6\kel$ stable warm absorber phases and the sixth column the
range of $\log(\xi/T)$ over which a multiphase medium is obtained.}
\labtablem{table1}
\end{center}
\end{table}

A detailed comparison between the stable phases
for warm absorbers predicted by
the two different
versions of CLOUDY is done in \tablem{table1}.
The second column of the table shows that for an absorber at $T\sim10^5\kel$,
C84 predicts $\xi_5\sim45$,
as compared to $\xi_5\sim74$ obtained
from the C07 stability curve, where $\xi_5$ is defined to
be the ionization parameter corresponding to the middle of
the $10^5\kel$ stable warm absorber phase.
Both C84 and C07 predict two discrete phases of warm absorber at
$\sim10^5$ and $\sim10^{5.7}\kel$ which are in pressure equilibrium
with each other and have been highlighted in \fig{stability}.
The C07 curve continues to have stable thermal states at
$\sim 10^6\kel$ which is not true in the C84 case.
The range of $\log(\xi/T)$, over which the warm absorber exists are
given in the fourth and fifth columns of respectively for the low and high
temperature states. The extent of the $10^5\kel$ phase in $\log(\xi/T)$ is
about four times larger in C07, predicting greater probability of
finding $10^5\kel$ warm absorbers. In the sixth column we have compared
the range $\Delta_{\rm{M}}\log(\xi/T)$ where the warm
absorber exhibits multiple phases. C07 predicts a 40\% larger
range and hence greater possibility of a multiphase warm absorber.

In order to isolate the atomic physics underlying the change in the
stability curve, we have plotted the fractional variation of temperature
$\Delta T/T_{C07} = (T_{C84}-T_{C07})/T_{C07}$, from one version to
another, against $\log \xi$ in the top panel of \fig{compare}.
We see that $\Delta T > 0$ for a major part of the range
$1.0 < \log\xi < 4.5$, so that the gas is predicted to be cooler by C07.
The cooling fractions ($\Delta C$) of the
major cooling agents and the
heating fractions ($\Delta H$) contributed by the principal heating
agents using C07 are plotted against $\log \xi$ respectively in the
middle and bottom panel of \fig{compare}. The ions that contribute
significantly where $\Delta T/T_{C07} \gtrsim 0.5$, are He$^{+1}$ and 
high-ionization species of silicon (+10 and +11) and iron
(+21, +22 and +23). In the same $\log\xi$ range, the principal heating
agents are highly ionized species of oxygen (+6 and +7) and iron (+17 to +25).

\begin{figure*}
\centerline{\psfig{figure=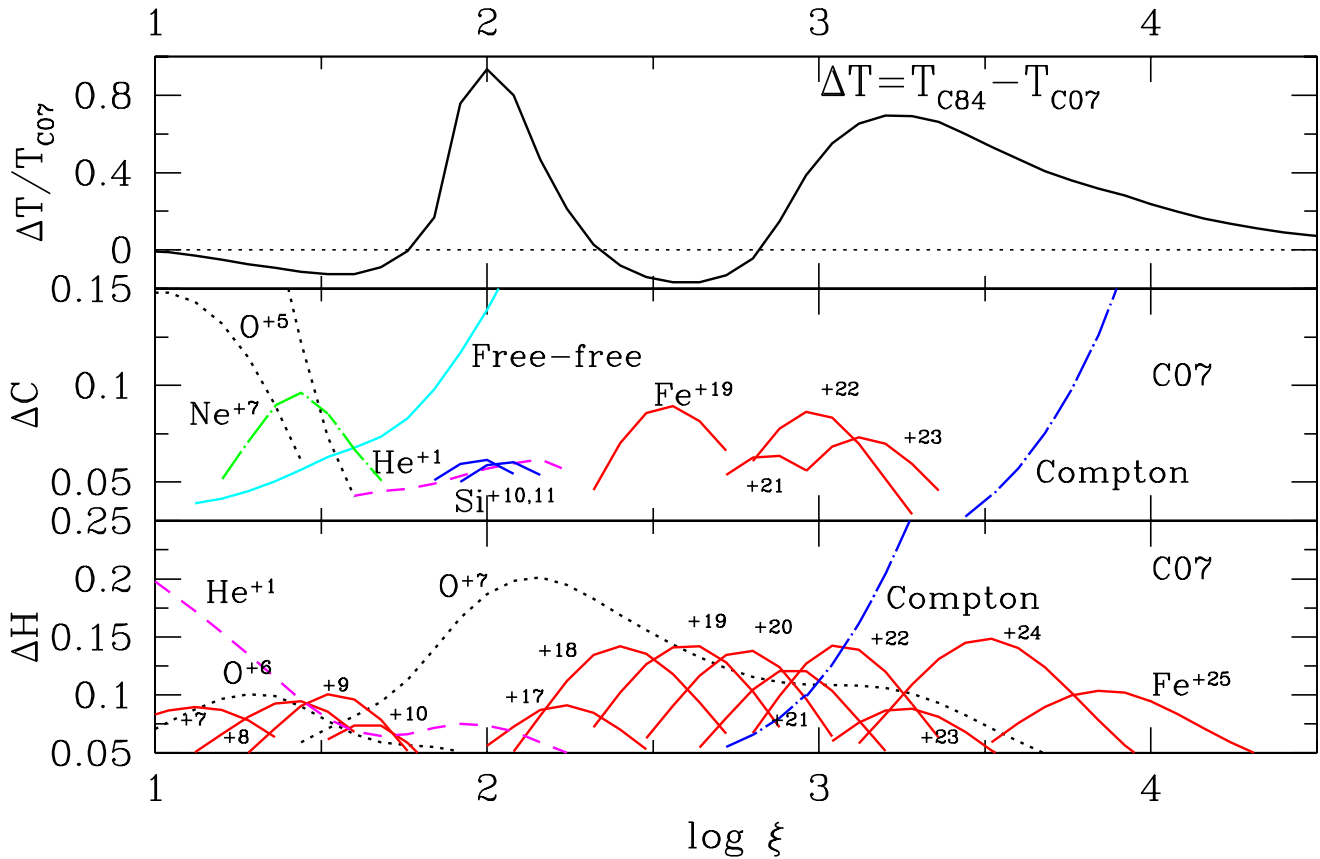,width=18cm,angle=0.0,bbllx=19bp,%
bblly=347bp,bburx=580bp,bbury=719bp,clip=true}}
\caption{Top : The fractional change  of temperature between the two versions
C84 and C07 of CLOUDY as a function of the ionization parameter. Middle : The cooling
fraction ($\Delta C$) for the major cooling agents as a function of
$\log\xi$, using C07. He$^{+1}$ and highly ionized species of silicon and
iron play important roles. Bottom : The heating fraction ($\Delta H$)
contributed by the principle heating agents as a function of the ionization
parameter, obtained using C07. Highly ionized species of oxygen like
O$^{+6}$, O$^{+7}$, and various ions of iron like Fe$^{+7}$ to Fe$^{+10}$ and
Fe$^{+17}$ to Fe$^{+25}$ are the key ingredients that are responsible for
heating the gas in the range $1.0<\log \xi<4.5$.}
\labfig{compare}
\end{figure*}


To identify the ions which are responsible for $\Delta T/T_{C07}
\gtrsim 50\%$, we compare their column densities predicted by C84
and C07. In the previous photoionisation calculations in this
paper it was sufficient to use one zone models for optically thin
gas. However, to calculate and compare the column densities of
the ions over the range $1.0<\log\xi<4.5$, we chose to specify
the gas to have total hydrogen column density
$\nh=\onlyten{22}\cmsqi$, which is typical for warm absorbers.
The column densities are plotted in \fig{colmdnsty} with solid
lines for C07 and dashed lines for C84. It is seen that the
column densities of the major coolants changed significantly. The
cooling agents are among the ions for which dielectronic
recombination rate coefficients (hereafter DRRC) have been
updated to the references below, as will be discussed in detail
in \secn{drrc}. Thus the enhanced cooling in C07 due to the
change in DRRC is the cause of the shift in the stability curves.


\section{ Changes in Dielectronic Recombination Rate Coefficients}
\labsecn{drrc}


\begin{table}
\begin{center}
\begin{tabular}{c c c c c }
\hline
& & & \multicolumn{2}{c}{Recombination rates ($\rm{s}^{-1}$)} \\ \cline{4-5}
\raisebox{1.5ex}[0cm][0cm]{Ion} & \raisebox{1.5ex}[0cm][0cm]{$\log T$} & %
\raisebox{1.5ex}[0cm][0cm]{$\log \xi$} & \raisebox{-0.5ex}[0cm][0cm]{C07} & %
\raisebox{-0.5ex}[0cm][0cm]{C84} \\
\hline \\
He$^{+1}$ & 5.34 & 2.00 & 1.66 & - \\
Si$^{+10}$ & " & " & 1.36 $\times 10^{-1}$ & 2.26 $\times 10^{-2}$ \\
Si$^{+11}$ & " & " & 8.85 $\times 10^{-2}$ & 2.50 $\times 10^{-2}$ \\ \\ \hline \\
Fe$^{+21}$ & " & " & 9.06 $\times 10^{-2}$ & 2.20 $\times 10^{-2}$ \\
Fe$^{+22}$ & " & " & 8.36 $\times 10^{-2}$ & 2.37 $\times 10^{-2}$ \\
Fe$^{+23}$ & " & " & 5.77 $\times 10^{-2}$ & 2.28 $\times 10^{-2}$ \\
\hline
\end{tabular}
\caption{Recombination rates of the dominant cooling agents
obtained using C84 and C07. The ions are given in column 1. The
temperature and ionization parameter at which the recombination
rates have been obtained are noted in columns 2 and 3 respectively.
The fourth and the fifth columns respectively give the value of the
recombination rates predicted by C07 and C84.}
\labtablem{table2}
\end{center}
\end{table}



\begin{figure*}
\centerline{\psfig{figure=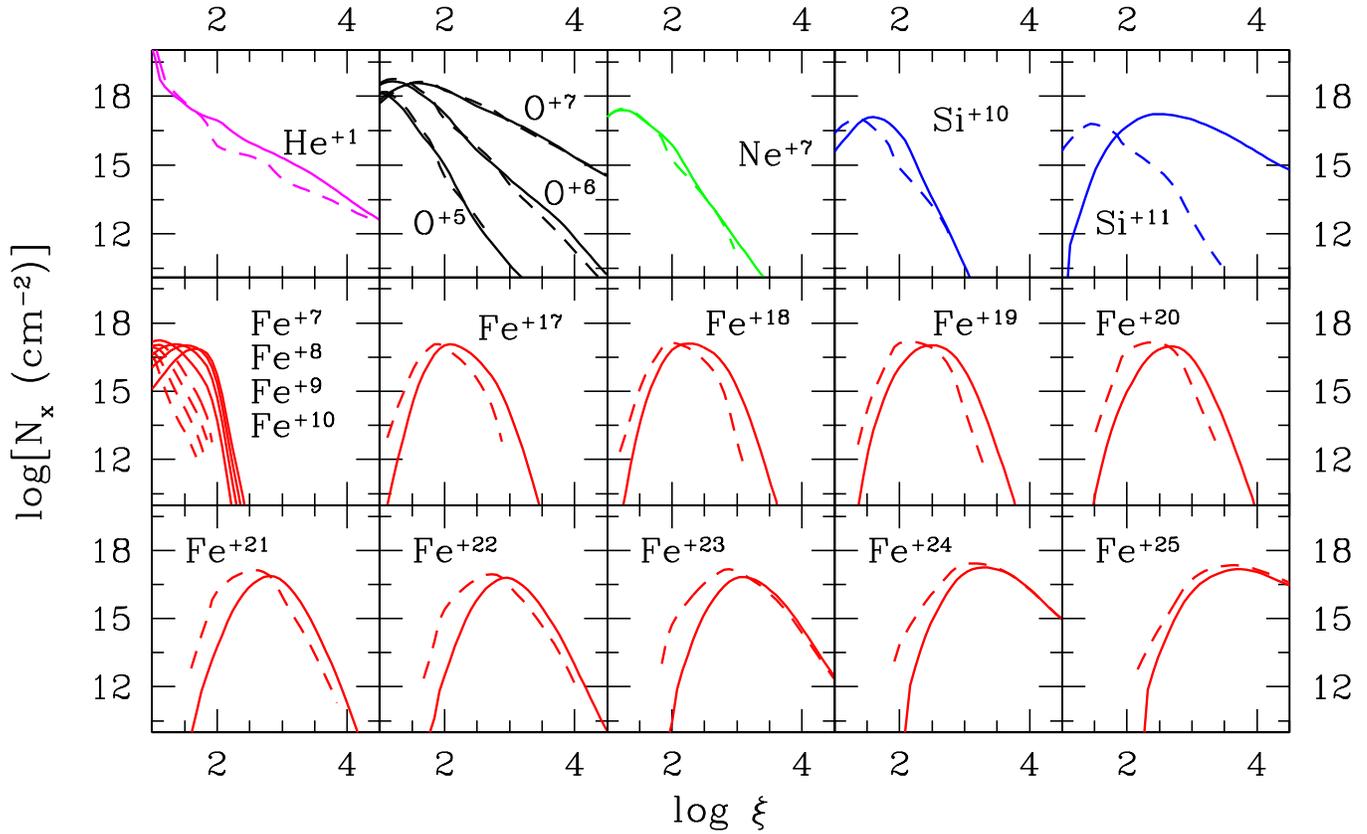,width=18cm,angle=0.0,bbllx=19bp,%
bblly=144bp,bburx=592bp,bbury=530bp,clip=true}}
\caption{The column densities for the same ions as in the middle and bottom panel
of \fig{compare}. The solid lines
are predicted by C07 and the dashed lines are for C84. The difference in
the column densities are more pronounced for the ions which are
significant cooling agents, with Si$^{+11}$ showing as much as a difference
of about two orders of magnitude at $\log\xi\sim2.0$ where it acts as the
dominant cooling ion.}
\labfig{colmdnsty}
\end{figure*}


The evolution of the thermal phases from the nebular temperatures
of $10^4\kel$ to the coronal temperatures of $10^6\kel$ depends sensitively
on the detailed atomic physics of the various elements which
contribute to photoelectric heating as well as to cooling due to recombination
and collisionally excited lines. In \tablem{table2}, we have compared the
total recombination rates (dielectronic + radiative)
for the significant cooling agents as predicted by C07 and C84.
The values for $\log T$ and $\log \xi$ at which the
comparisons have been made are given in columns 2 and 3 respectively.
For all the ions the total recombination rates are
significantly higher in C07 than in C84 as shown by columns 4 and 5.
Referring to \fig{compare}, we see that the differences in total
recombination rates are relatively larger for ions like Si$^{+10}$,
Si$^{+11}$, Fe$^{+21}$ and Fe$^{+22}$ which are significant
cooling agents for $\log \xi \sim 2.0$ and $\log \xi \sim 3.2$,
corresponding to which we have maximum difference in predicted equilibrium
temperatures $T_{C07}$ and $T_{C84}$.

In the warm absorber temperature range $10^5 \lesssim T \lesssim
10^7\kel$, dielectronic recombination dominates over radiative
recombination for many ions (Osterbrock \& Ferland, 2006). Unlike
the radiative recombination rate coefficients, the DRRC have
undergone significant changes over the last decade. C84 used DRRC
from Nussbaumer \& Storey (1983;1984; 1986; and 1987) and Arnaud
\& Raymond (1992). In C07 we have taken the DRRC for the
isoelectronic sequences of lithium, beryllium, boron, carbon,
nitrogen, oxygen and iron-like ions respectively from Colgan
\etal (2004), Colgan \etal (2003), Altun \etal (2004), Zatsarinny
\etal (2004a), Mitnik \& Badnell (2004), Zatsarinny \etal (2003),
and Gu (2003). The DRRC for Ne to Na like ions and Na to Mg like
ions are taken from Zatsarinny \etal (2004b) and Gu (2004). The
C07 DRRC for any given ion is usually substantially larger than
the C84 DRRC when the temperature is much lower than the
ionization potential of the ion. The significant cooling agents
in C07 are among these ions. This indicates that the updated DRRC
database in C07 is the cause of the changes in the stability
curve.

It is seen from \tablem{table2} that the increase in recombination
rates from C84 to C07 is larger for the lower ionization species.
The reason for the large increase, in general, in the
low-temperature DRRC (referenced above) is the explicit inclusion
of the contribution from low-lying (in energy) level-resolved
dielectronic recombination resonances which have been accurately
positioned by reference to the observed core energies. Unlike
high-temperature dielectronic recombination, where the full
Rydberg series contributes, low-temperature dielectronic
recombination is not amenable to simple scaling or empirical
formulae or guesses. Zatsarinny et al (2003) illustrate the
radical and erratic changes in the low-temperature DRRC on moving
between adjacent ions of the same isoelectronic sequence and,
even, on changing from a term-resolved picture to level-resolved
one for the same ion.

The DRRC database is still not complete, specially for the lower
ionization states. For ions which do not have computed DRRC
values, C07 uses a solution, as suggested by Ali \etal (1991):
for any given kinetic temperature, ions that lack data are given
DRRC values that are the averages of all ions with the same
charge. The advantage of this method is that the assumed rates
are within the range of existing published rates at the given
kinetic temperature and hence cannot be drastically off. However
we have checked whether the points in the C07 stability curves
have computed DRRC values or guessed average values,
concentrating on the parts of the curve which have multiphase
solutions for the warm absorber ($10^5 \lesssim T \lesssim
10^6\kel$) and are different from the C84 stability curve. We
find that all the ions which act as major cooling agents for each
of these points in the stability curve have reliable computed
DRRC values. Thus, the new data base provides a more robust
measurement of the various physical parameters involved in
studying the thermal and ionization equilibrium of photoionised
gas.


\section{Conclusion}
\labsecn{conclusion}

We have shown that stability curves for warm absorbers in AGN
generated by two versions of the ionization code CLOUDY, C84 and
C07, for the same physical conditions, are substantially
different in shape, leading to different conclusions regarding
the nature of the warm absorber. The differences in the results
of the photoionisation calculations arise due to major changes in
the dielectronic recombination rate coefficient (DRRC) data bases
which have taken place over the last decade. The modern version
C07 includes reliable computed DRRC values for many more ions
than C84 for the entire part of the stability curve relevant for
warm absorbers and does not rely on guessed average values. Thus,
the physical nature of the warm absorber predicted by modern
calculations are more reliable. We suggest that past calculations
for photoionised plasma in thermal equilibrium should be
reconsidered, taking into account these changes. Since, any
other atomic physics code (eg. XSTAR), in popular use for warm
absorber studies, is also likely to be affected by the changes in
the DRRC data base, caution should be exercised while using or
comparing results from them as well.

\section*{acknowledgements}
We thank Smita Mathur for reading an early version of the draft and providing helpful suggestions.

\end{document}